\pdfoutput=1
\documentclass[aps,prb,onecolumn,showpacs,preprintnumbers,amsmath,amssymb,groupedaddress]{revtex4-1}
\usepackage{mathptm} 
\usepackage{dcolumn}                    
\usepackage{bm}                        
\usepackage{graphicx}
\usepackage{times}
\begin{document}
\newcommand {\be}{\begin{equation}}
\newcommand {\ee}{\end{equation}}
\newcommand {\bea}{\begin{eqnarray}}
\newcommand {\eea}{\end{eqnarray}}
\newcommand {\nn}{\nonumber}
\newcommand {\bb}{\bibitem}
\newcommand{\et}{{\,\it et al\,\,}}
\newcommand{\CuB}{CuBiS$_{2}$}
\newcommand{\BiTe}{Bi$_{2}$Te$_{3}$ }

\title{Acoustic impedance and interface phonon scattering in Bi$_2$Te$_3$ and other semiconducting materials}

\author{Xin Chen, David Parker and David J. Singh}
\address{Oak Ridge National Laboratory, 1 Bethel Valley Rd., Oak Ridge, TN 37831}

\date{\today}

\begin{abstract}
We present first principles calculations of the phonon dispersions of \BiTe and discuss these in relation to the acoustic phonon
interface scattering in ceramics.  The phonon dispersions show agreement with what is known from neutron scattering for the optic modes.
We find a difference between the generalized gradient approximation and local density results for the acoustic branches.  This is a 
consequence of an artificial compression of the van der Waals bonded gaps in the \BiTe structure when using the generalized gradient approximation.  
As a result local density approximation calculations provide a better description of the phonon dispersions in Bi$_{2}$Te$_{3}$.  A key characteristic 
of the acoustic dispersions is the existence of a strong anisotropy in the velocities.  We develop a model for interface
scattering in ceramics with acoustic wave anisotropy and apply this to \BiTe and compare with PbTe and diamond.

\end{abstract}
\maketitle
\section{Introduction} 

Thermoelectric performance is commonly quantified in terms of a dimensionless parameter ZT, defined as follows:
\bea
ZT &=& \frac{S^2\sigma T}{\kappa}
\eea
Here $S$ is the Seebeck coefficient, $\sigma$ the electrical conductivity, and $\kappa$ the thermal conductivity.  It is usually a good approximation
to treat $\kappa$ as being comprised of a lattice portion and an electronic portion.  The electronic portion is directly related to the electrical conductivity by the Wiedemann-Franz relation
(usually a good approximation for the heavily doped semiconductors that are useful thermoelectrics), leading to the point that, from a standpoint of materials optimization,
the lattice thermal conductivity represents wasted heat transfer and should be as small as possible.  One realization of this is the ``phonon glass electronic crystal" (PGEC)  concept of Slack \cite{slack}, 
in which phonons are strongly scattered, leading to low lattice thermal conductivity, while the charge carriers are not strongly scattered.  The filled skutterudites
\cite{sales, subramanian,shi} represent an apparent realization of the PGEC concept.  Another approach towards the PGEC concept is the use of nanostructuring from compaction and sintering of a nanosize powder into a ceramic.  The mean free path for phonons in bulk crystalline thermoelectrics is often one to two orders of magnitude larger than that for electrons, so the use of grain sizes in between these two mean free paths will tend to strongly scatter phonons, but not electrons.  This approach has been successfully 
applied to Bi$_2$Te$_{3}$ \cite{poudel,tritt}, raising the ZT values near room temperature from the previously found value of 1.0 to an impressive 1.5.  A similar scenario could also apply to hole-doped Bi$_2$Se$_3$\cite{parker}.  Note that there are various types of grain boundary scattering that can reduce the thermal conductivity, such as insulating interstitial material, but ytpically these destroy the electrical conductivity, preventing thermoelectric performance.  In this respect, the sintering of polycrystalline samples (as opposed to simply compressed powder) is important as it assures good electrical contact between the nanograins, while maintaining interface scattering.  Understanding interface phonon scattering at such electrically conducting interfaces is therefore of importance.  

One factor as yet unaddressed, however, is the role of phononic anisotropy in producing phononic scattering.  Consider by analogy, for example, the case of light propagation in dense ceramics.  In that case it is known that fine-grained ceramics of optically isotropic materials can be made transparent \cite{jiang}, while this is not the case for anisotropic materials with random grain orientation.  In the former case, despite the small grain sizes, light is able to pass through the material because there is not significant scattering at the grain boundaries.  The reason for this is that the light speed does not change at the grain boundary, so there is no impedance,or velocity mismatch.   While the acoustic case differs from the optical case due to the presence of the longitudinal mode, the velocity mismatch still applies.   

We note also that, unlike in optics where there is no optical anisotropy in cubic materials, a cubic material can have a rather anisotropic elastic response tensor (the tensor of elasticity C$_{ijkl}$) and hence sound speed - a good example of such a material is PbTe, as described in more detail in the last section.  The reason for this is that the response tensor for optics - the dielectric constant tensor - is a {\it second} rank tensor whose off-diagonal elements must necessarily vanish due to the cubic symmetry.  However, no such restriction applies to the {\it fourth} rank-tensor of elasticity, and indeed the sound velocity in a cubic material can vary significantly by direction, as will be evident for PbTe.

Note that for heat transport in crystalline solids it is the longer wavelength acoustic modes that dominate heat transport due to the presence of a significant group velocity - the sound speed.  The main point of this paper is that it is this sound velocity mismatch that is ultimately responsible for the efficacy of grain boundary scattering, and that this mismatch can be quantitatively assessed by considering the sound velocity anisotropy.  We provide a definition, and examples of, simple dimensionless parameters $R$ and $S$, easily computable if the elastic constants are known, that should provide valuable information about the ability of nanostructuring (of a type that yields grains in intimate contact) to reduce lattice thermal conductivity.  

We will make direct application of our findings to Bi$_2$Te$_3$, which as already mentioned has already shown substantial performance benefits from nanostructuring, and in addition PbTe.  One conclusion to be drawn from our work is that, unlike in electronic transport, where anisotropy is generally destructive to thermoelectric performance, in phononic transport anisotropy is beneficial by enhancing the effects of nanostructuring in reducing the lattice thermal conductivity.

\section{Phonon calculations for Bi$_2$Te$_3$}

With an eye towards the effects of nanostructuring in reducing $\kappa_{lattice}$ and enhancing ultimate performance we have computed the phonon dispersions and density-of-states for Bi$_2$Te$_3$, the best known and most studied thermoelectric.  Our calculations are based upon density functional theory in the framework of Bl\"ochl's projector augmented-wave (PAW) method \cite{a} within the 
local density approximation (LDA) as implemented in VASP \cite{vasp}.   We also did gemeralized gradient approximation calculations but found that they are not as accurate as the LDa results (see below).  A 3x3x3 $k$-point grid in a 3x3x3 supercell was used, along with an energy cutoff of 300 eV.  Cell parameters and internal coordinates were both relaxed until internal forces were less than 2 meV/\AA.  From the computed electronic structure one performs several supercell calculations incorporating ``frozen-phonons", or atomic displacements dictated by the rhombohedral crystal symmetry.  By evaluating the forces on the displaced atoms one may generate a basis set of force constants from which the phonon band structure and density-of-states are generated.  We depict these in Figure 1.  Previous Bi$_2$Te$_{3}$ lattice dynamics calculations were performed in Refs. \onlinecite{qiu,huang,wang,kullmann}.  Spin orbit coupling was not included and we therefore cannot assess the claim of Ref. \onlinecite{wang} for evidence of a spin-orbit coupling-related lattice instability.  Experimentally the material is known to be stable.  Our calculations generally reproduce the non spin-orbit coupling phonon dispersions of these authors.

One notes upon examination of the central region (the portions Z-$\Gamma$ and $\Gamma$-L) that the three acoustic modes differ significantly in these two directions (respectively c-axis and in-plane).  In particular, the highest velocity acoustic mode, the longitudinal acoustic, has significantly lower velocity in the c-axis direction $\Gamma-Z$ than the nearly planar direction $\Gamma-L$; quantitatively, the c-axis longitudinal velocity is 1811 m/sec and the planar is 2394 m/sec, a difference of about 30 percent. In addition, the transverse acoustic modes are degenerate from $\Gamma$-Z but not so in plane; here the velocities are significantly different as well, with the single c-axis value of 1774 m/sec and the two planar velocities of 1395 and 1728 m/sec.  These velocities are low and generally typical of good thermoelectric materials.  Note also that in the frequency range at and above 1 THZ the optic modes intersect with the acoustic modes, so that the region of heat transport is limited to less than 1 THZ, which in turn limits the phonon momenta that contribute to transport to locations relatively near the $\Gamma$ point.

Turning to the phonon density of states, one finds three regions of interest.  Highest in frequency, as expected given the lighter mass, are the primarily Te optic modes between 2.3 and 4 THZ.  As noted previously, these contribute little or nothing to the phononic transport due to the very small group velocities (see left hand panel of Fig. 1).  A similar statement applies to the primarily Bi optic modes between 
1 and 2.3 THZ ,although these may be important contributors to phononic scattering due to anharmonic scattering of the lower frequency acoustic modes, which are at frequencies less than 1.5 THZ.  As noted above,  only a fraction of these modes - those less than 1 THZ  - contribute to thermal transport as the higher frequency acoustic modes are strongly scattered by the adjacent optic modes, and also have smaller group velocities.  For example, in the $\Gamma-L$ direction only those acoustic phonons less than half the L-point momentum will contribute to heat transport, while in the $\Gamma-F$ direction this cut-off frequency occurs at a momentum roughly sixty percent of the F-point momentum.

The original lattice dynamics calculations for this work employed the standard GGA \cite{b,c}.  However, these results produced longitudinal sound speed velocities which were higher in the $\Gamma$-Z direction (the c-axis) than in the planar directions.  This result persisted even when a relatively fine 4x4x4 $k$-point mesh was used.  Similarly, we initially found from  first principles calculations of the elastic constants of Bi$_2$Te$_3$ using WIEN2K  \cite{wien} and the GGA that the elastic constant c$_{11} < c_{33}$, which implies lower longitudinal sound speeds in plane.  All these results are contrary to the elastic constant data of Jenkins \cite{jenkins}, which produces higher longitudinal sound speeds in-plane (see the next section), as well as the measured planar and c-axis thermal conductivity \cite{spitzer}, where the c-axis value is less than half the planar value, indicating lower c-axis sound speeds.   Our calculational discrepancy was likely due to the common GGA overestimation of lattice constants.   It was for this reason that we performed these lattice dynamics calculations within the LDA which often gives structural and elastic properties in better agreement with experiment \cite{fast,mehl}. These results suggest that for anisotropic layered semiconductors such as Bi$_2$Te$_3$, use of the LDA to compute elastic and lattice dynamics properties may be desirable.

\begin{figure}[h!]
\includegraphics[width=6.5cm,angle=270]{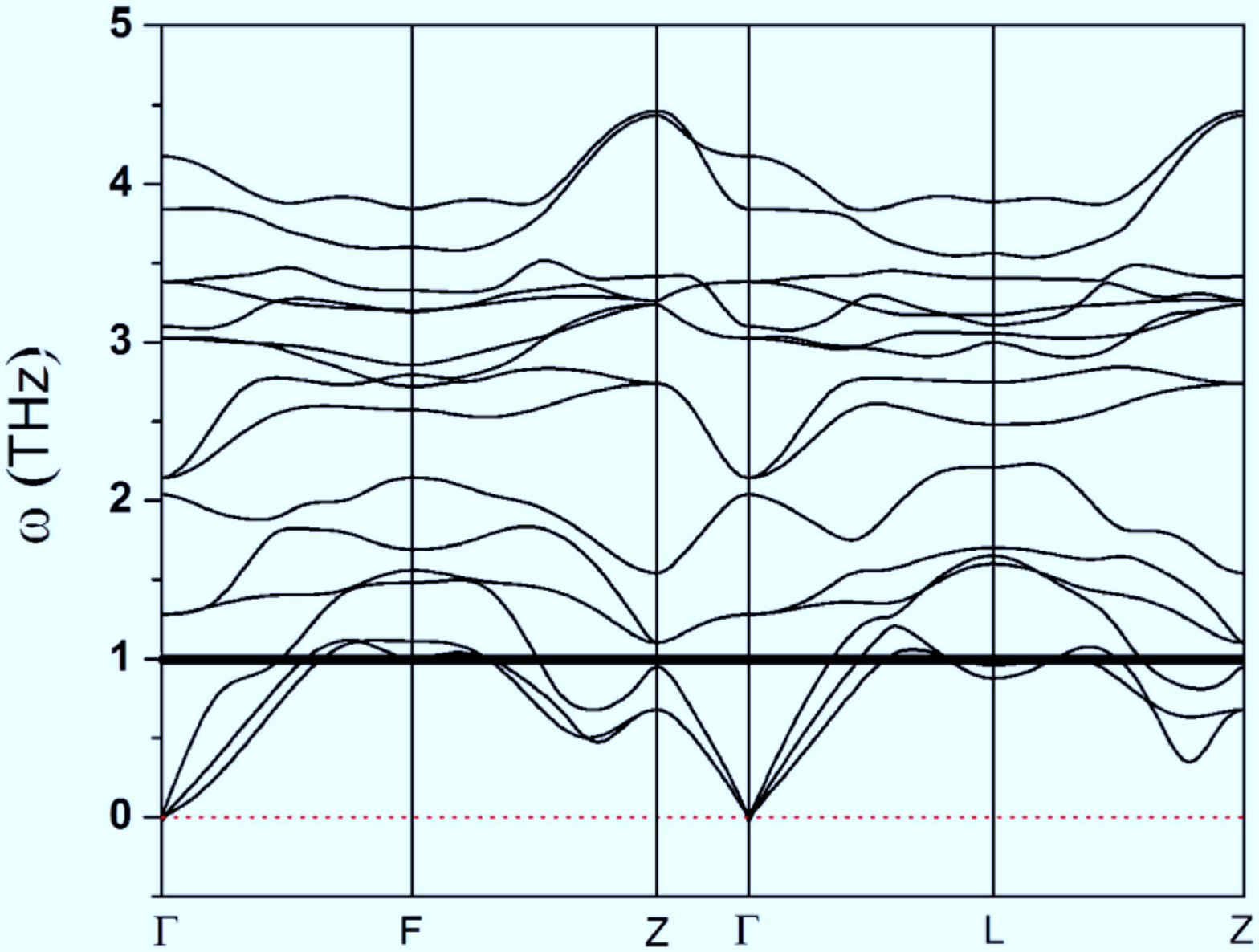}
\includegraphics[width=7cm,angle=270]{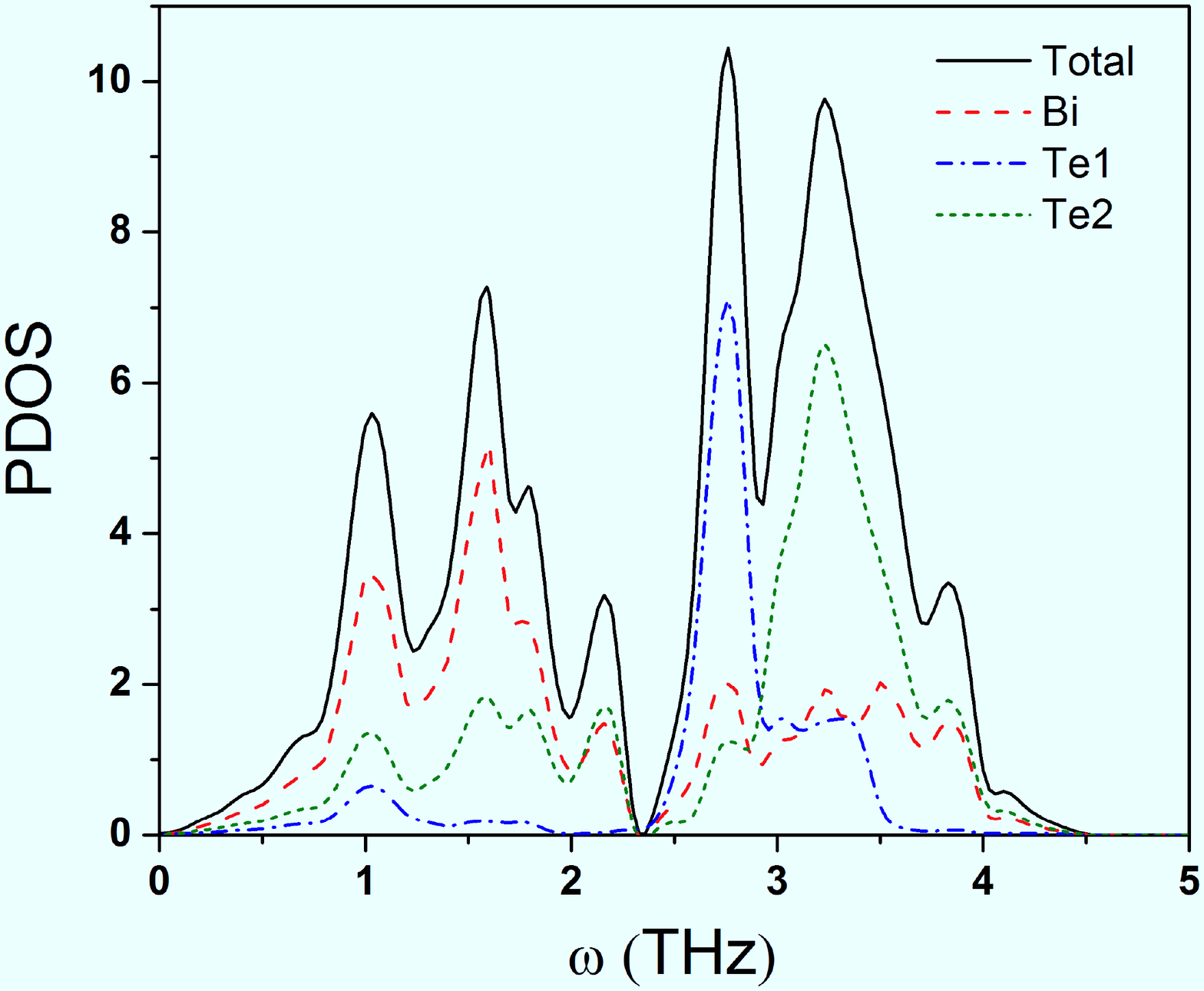}
\caption{The computed phonon dispersions (left) and associated density-of-states (right)  of Bi$_2$Te$_3$.  The heavy black line in the left figure indicates the upper frequency limit of the heat carrying acoustic modes.}
\end{figure}

\section{Sound speed anisotropy and nanostructuring effectiveness parameters}

The discussion of the previous section makes plain the significant anisotropy in the phonon transport of Bi$_2$Te$_3$.  Here we find a way to assess the quantitative impact of this anisotropy in reducing lattice thermal conductivity and make a comparison with another well-studied thermoelectric material, PbTe, as well as diamond.

Our method is the following.  From published calculated values of the elastic constants for Bi$_2$Te$_3$ one may generate the associated Christoffel \cite{musgrave} elastic tensor stiffnesses and solve the resulting secular equation for the three sound velocities (one longitudinal and two transverse) as a function of wave propagation direction.  The relevant equations may be found in Ref \onlinecite{mayer}.  We have plotted up the sound speeds, as a function of propagation direction, in Figure 2.  The sound speed plots differ significantly from a spherical shape, underscoring the anisotropy already apparent from the calculated phonon bandstructure.

We turn now to the impact of the anisotropic sound speeds on the lattice thermal conductivity.  As is well known, the lattice thermal conductivity $\kappa_{l}$ is given as
\bea
\kappa_l &=& \sum_{{\bf q}, i} C_{{\bf q},i} v_{{\bf q},i} \ell_{{\bf q},i}
\eea
where $C_{{\bf q},i} $ is the specific heat attributable to a phononic mode with momentum {\bf q} and polarization $i$, $v$ the sound speed of that mode and $\ell$ the mean free path of that mode, and a sum over the modes of significant group velocity is taken.  
In general, the fraction R of elastic energy reflected at a grain boundary interface at normal incidence is given by the impedance mismatch formula:
\bea
R &=& \left(\frac{(Z_1-Z_2)}{(Z_1+Z_2)}\right)^2
\eea
Here Z$_{1}$ and Z$_{2}$ are the acoustic impedances of the two adjoining grains, given by Z$_{i}=\rho_{i} v_{i}$, where $\rho$ is the density of a grain and $v$ the sound speed (of a given polarization) within the grain.  Since we expect $\rho$ to be constant within a nanostructured sample, the energy fraction reflected depends on the sound speeds v$_{1}$ and v$_{2}$ in the two grains, {\it at the directions of incidence and transmission}, and in addition on the polarization of the incoming wave.  Note also \cite{ewing}  that an incident wave of one polarization may induce scattered waves of other polarizations, complicating the issue further.    Furthermore,  in a nanostructured sample we do not expect oriented grains.  Hence to work out the effective grain boundary scattering rate one must consider grain orientation as well as the intrinsic anisotropy of the sound speeds.  This becomes a rather difficult, and even difficult to formulate, problem when one realizes that the grains are not likely to be exactly randomly oriented, and that the degree of randomness will likely depend on the exact synthesis and nanostructuring techniques applied, unknown in this work.

Given that one purpose of this paper is to propose computationally simple nanostructuring effectiveness parameters, we therefore make a simple ansatz based upon the (relatively) random nature of the problem at hand.  Since the incident and transmitted velocities v$_{1}$ and v$_{2}$ are essentially uncorrelated, it is a fair approximation to replace v$_{2}$ in the above impedance mismatch expression by its average value (a similar assumption in a different context is made in many ``mean-field" theories) and integrate over all angles of incidence.  As with mean-field theories, the simpler expression is most quantitatively accurate when v$_{1}$ does not vary too much from its average.  The gross features of anisotropy, however, should be reasonably well captured by this expression.  The expression for R$_{total}$ is a simple two-dimensional integral:
\bea
R_{total} &=& \frac{1}{4\pi}\int  sin(\theta) d\theta d\phi \left(\frac{v(\theta,\phi)-v_{avg})}{(v(\theta,\phi)+v_{avg})}\right)^2
\eea
where the above integral is computed for each of the two transverse modes and the longitudinal mode and then averaged over the modes.  One could argue, based on phase space considerations, that the various terms should be weighted by the sound speeds, or sound speeds squared, or some other factor, of the various modes , but it is usually unclear in any given system what fraction of heat transport separately results from transverse and longitudinal modes \cite{footnote}, so we have retained the simplest possible expression.

The above expression yields a single number R$_{total}$ which gives in essence the average impedance mismatch reflected energy {\it at normal incidence} for a single scattering event.  It is typically fairly small - of the order of 0.01 or less even for highly anisotropic media, as depicted below.  However, there is an important additional scattering effect created by the velocity anisotropy.  As with propagation of electromagnetic waves, there is a form of Snell's law, $v_{incoming}/v_{transmitted}=\sin(\theta_{incoming})/\sin(\theta_{transmitted})$,  relating incoming and outgoing propagation angles (relative to the normal) to the relative sound speeds, and a version of total internal reflection, in which for certain angles of incidence there is {\it no} energy transmission across the interface, applies.  To put this quantitatively, for a 20 percent smaller sound speed (in a given direction) in the receiving material, angles of incidence greater than 53 degrees - {\it forty} percent of the possible angles of incidence  -  result in total internal reflection, even though the impedance mismatch reflection coefficient at normal incidence is only 0.012.  Since the {\it average} sound speeds in the two nanograins are of course equal, what one needs is a measure of the average deviation of the velocity from its average, which in essence is what the individual R$_{i}$ measures (more precisely speaking $R_{i} \simeq \frac{(\Delta v_{i})^2}{4v^{2}_{i,average}}$, where $\Delta v$ is the standard deviation of $v_i$ and v$_{average,i}$ the angular average sound velocity of a mode of polarization $i$.)

To give examples of velocity anisotropy and R$_{total}$ we have computed and present in Figures 2, 3 and 4 the sound speed anisotropy for three well known semiconducting materials: Bi$_2$Te$_3$, PbTe, and diamond, respectively.  Diamond is included to demonstrate a material with very low elastic anisotropy, while PbTe is another well known thermoelectric.  We will see that as expected, Bi$_2$Te$_{3}$ shows significant potential for nanostructuring reductions of $\kappa_{lattice}$.  As mentioned previously, despite its cubic structure, PbTe also shows large velocity anisotropy.  It might therefore also be expected to allow good reductions of lattice thermal conductivity due to nanostructuring, but recent work \cite{esfarjani}  shows that for PbTe these nanostructures must be smaller than 10 nm to have a significant effect, as the phonon mean free path is already very short.  Conversely, although diamond shows very low anisotropy, its phonon mean free path is so long (well over 100 nm) that small-grain nanostructuring would likely have a significant impact on its lattice thermal conductivity (noting nevertheless the impracticality of this material for thermoelectric applications).  We have taken account of this additional effect by defining a ``scattering potential coefficient" $S_{total}$ as R$_{total} \kappa_{lattice, bulk}/\kappa_{min}$, where $\kappa_{lattice,bulk}$ is self explanatory (values for Bi$_2$Te$_3$ and PbTe taken from Ref. \onlinecite{spitzer}) and $\kappa_{min}$ is the ``minimum thermal conductivity" \cite{slack2}, which we simply take as 0.5 W/m-K.
\begin{figure}[h!]
\includegraphics[width=5cm]{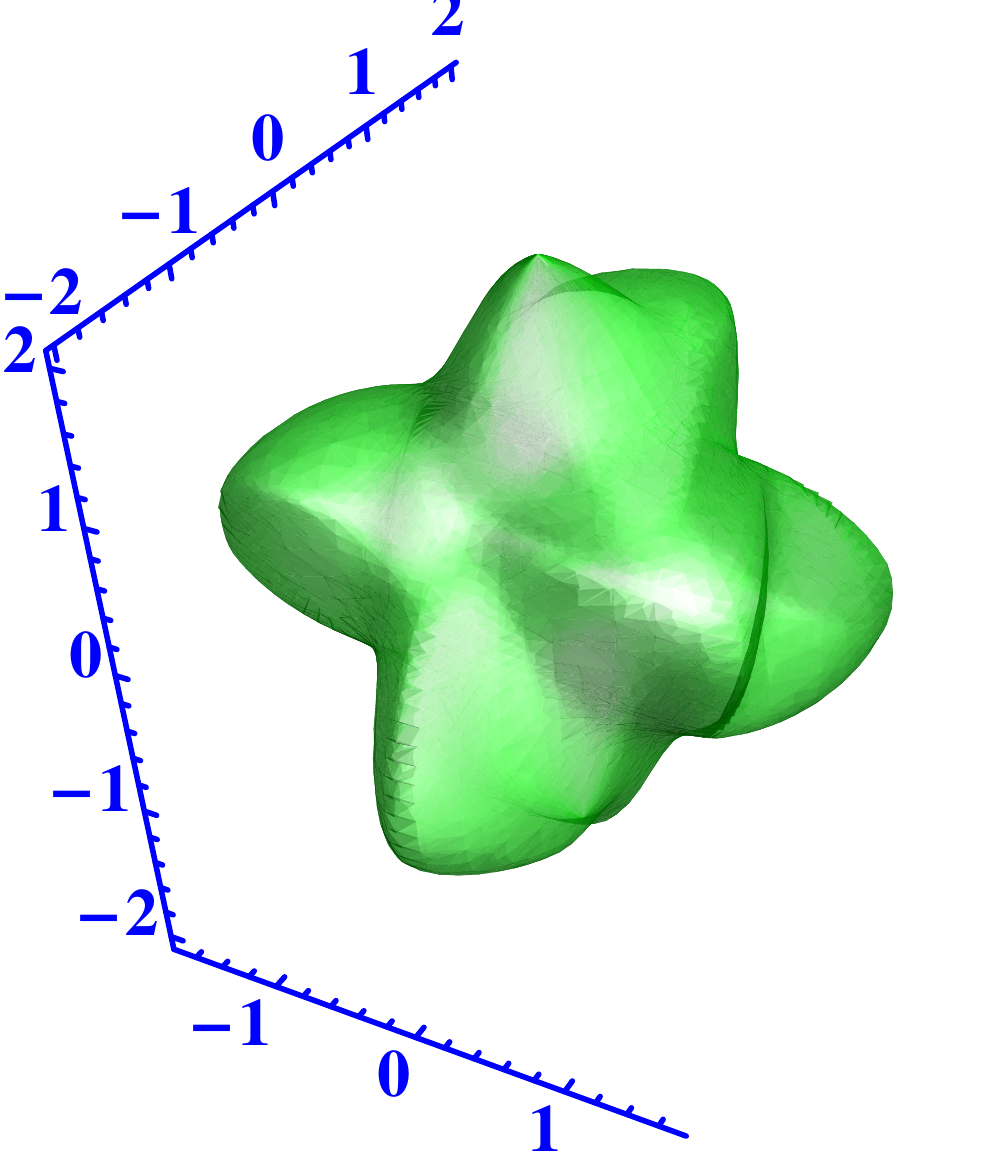}
\includegraphics[width=5cm]{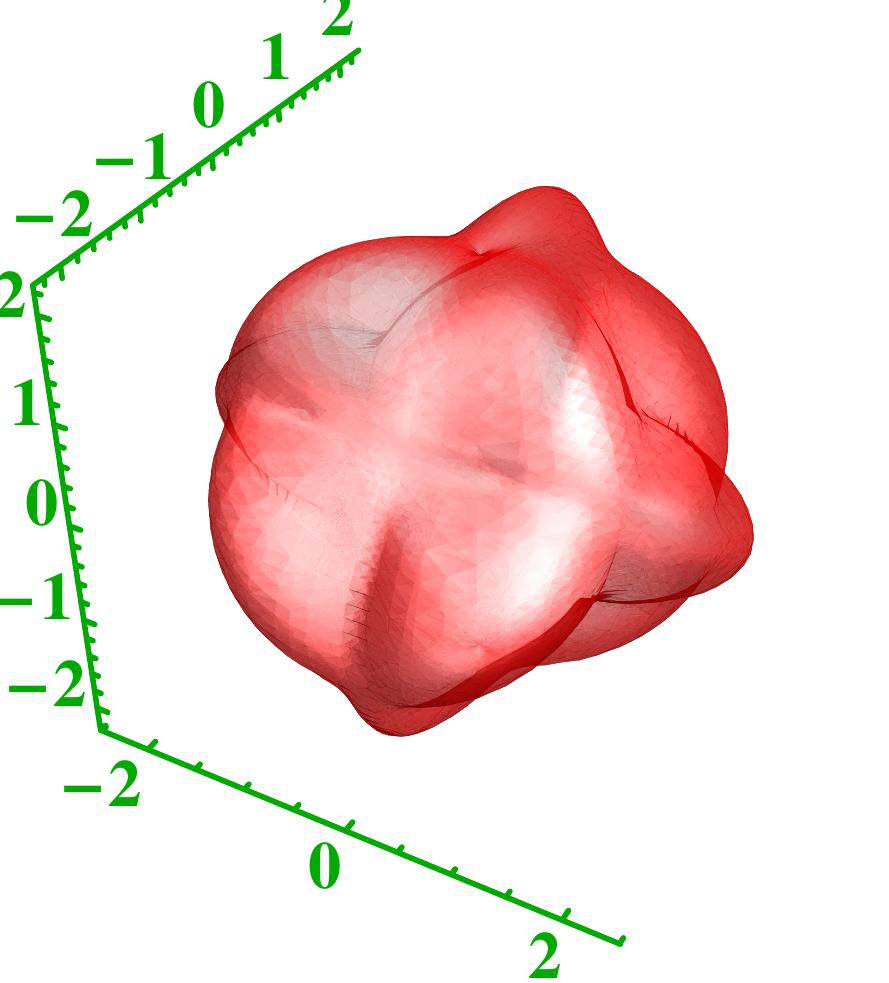}
\includegraphics[width=5cm]{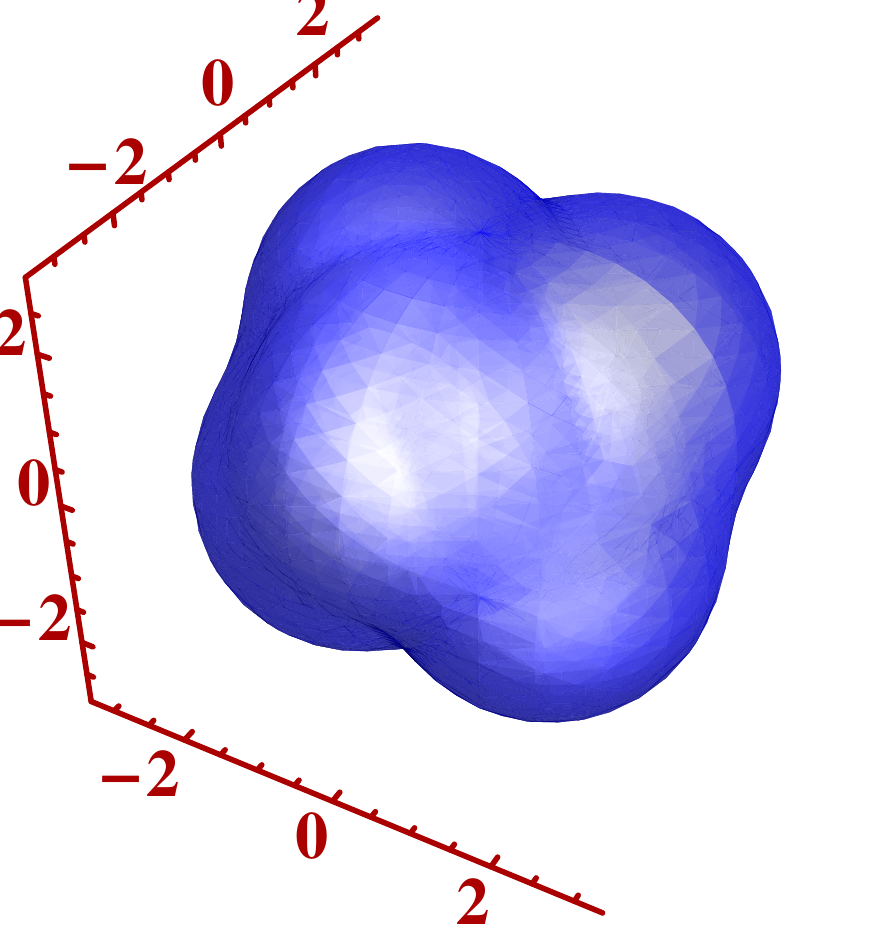}
\caption{The computed sound speed anisotropy of Bi$_2$Te$_3$.  Distance from origin represents sound speed, in km/sec, for that direction of propagation.  Transverse modes T1 and T2 left and center, respectively, longitudinal mode right.  Elastic constants taken from Ref. \onlinecite{jenkins}.}
\end{figure}

\begin{figure}[h!]
\includegraphics[width=5cm]{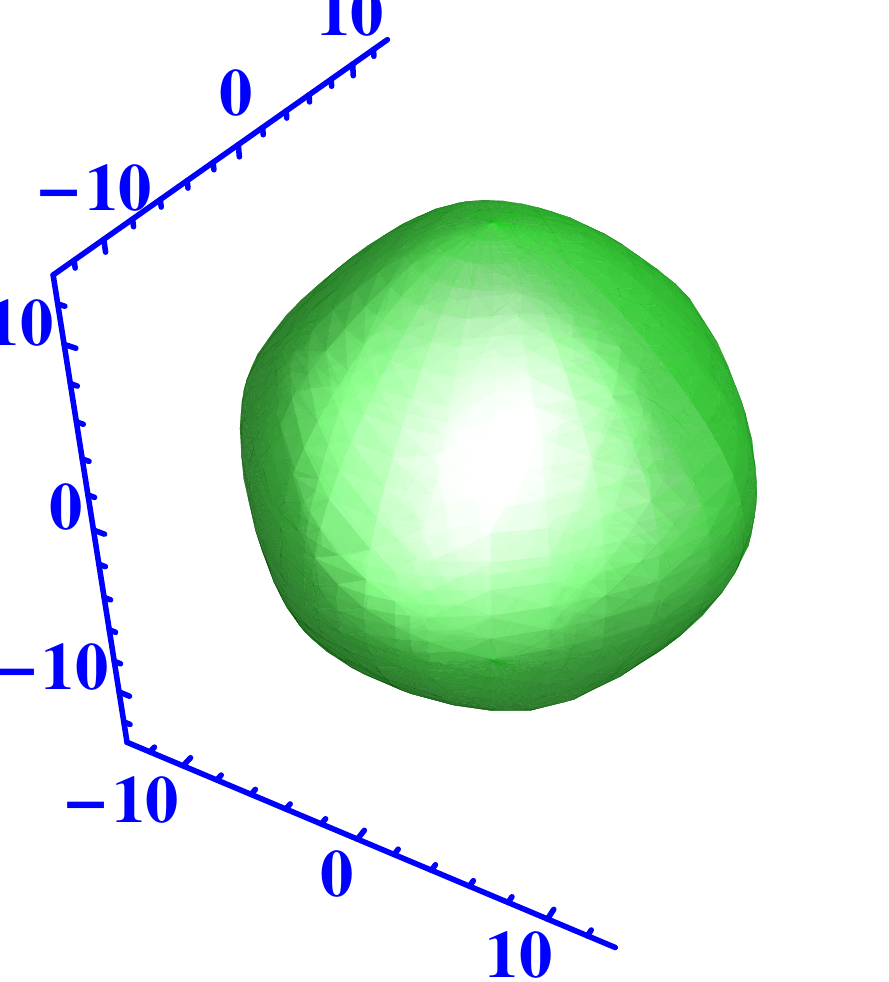}
\includegraphics[width=5cm]{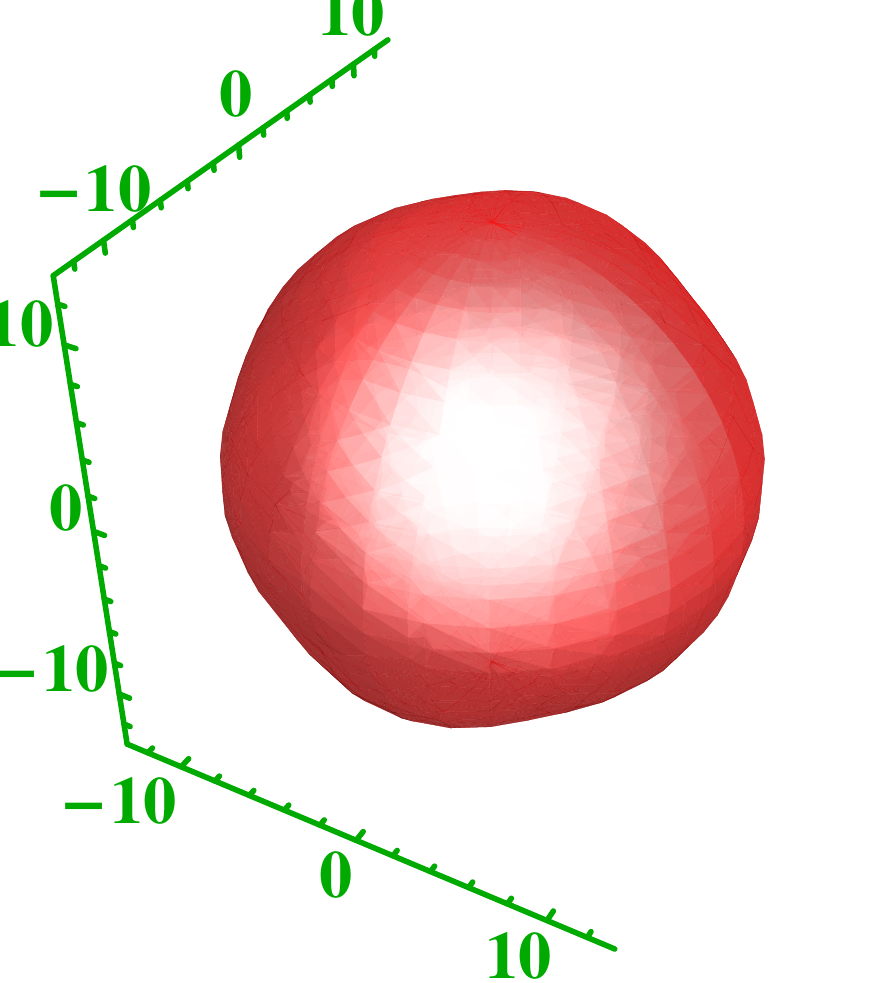}
\includegraphics[width=5cm]{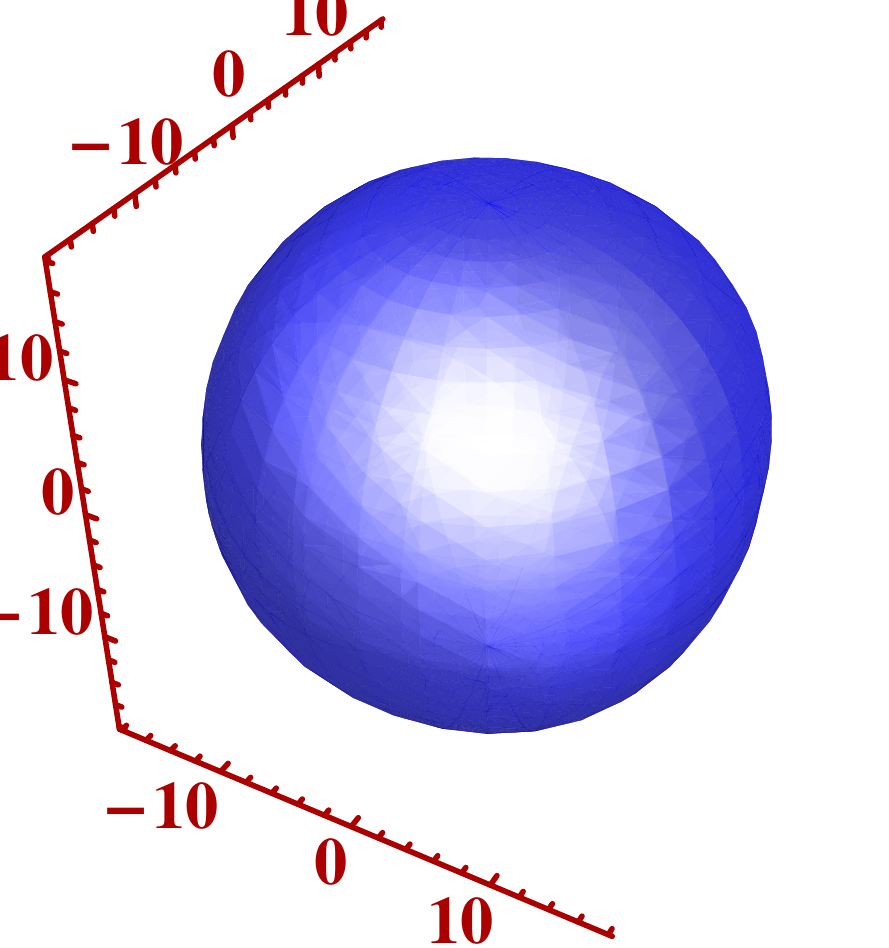}
\caption{The computed sound speed anisotropy of diamond.  Elastic constants taken from Ref. \onlinecite{grimsditch}.  Transverse modes left and center, longitudinal mode right.}
\end{figure}

\begin{figure}[h!]
\includegraphics[width=5cm]{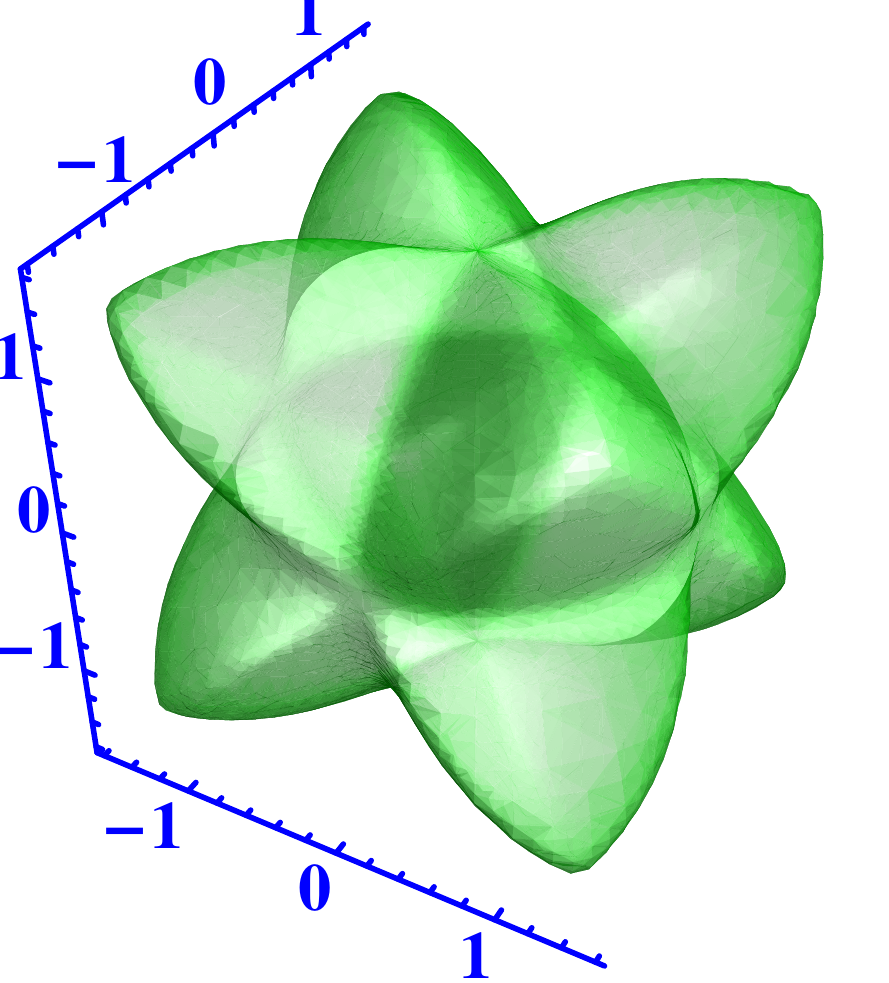}
\includegraphics[width=5cm]{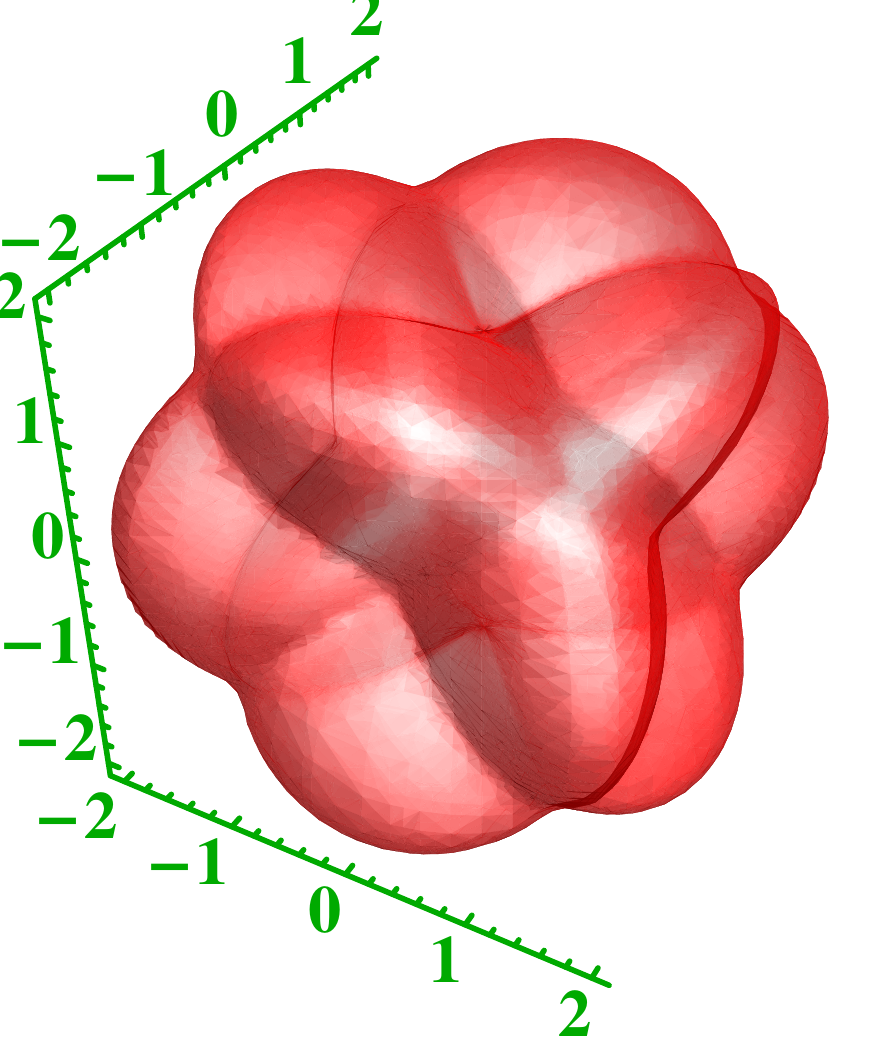}
\includegraphics[width=5cm]{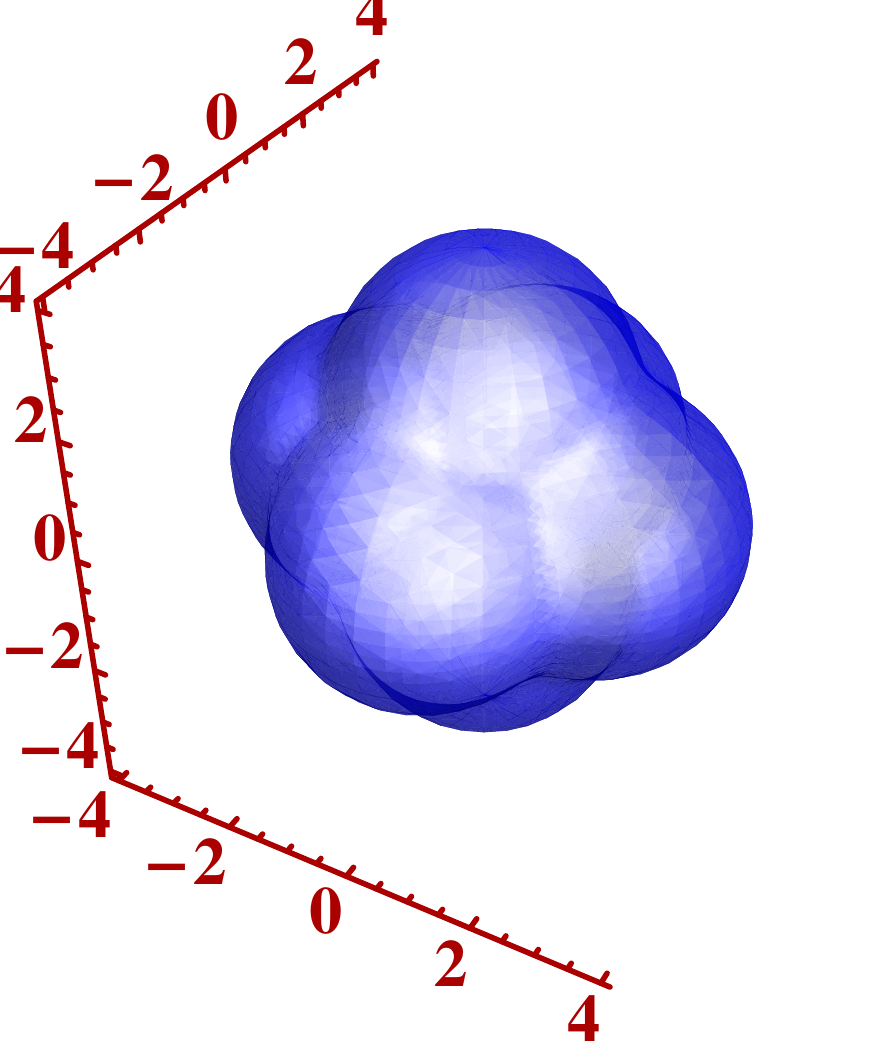}
\caption{The computed sound speed anisotropy of PbTe.   Elastic constants taken from Ref. \onlinecite{houston}.  Transverse modes left and center, longitudinal mode right.}
\end{figure}

In an attempt to quantify the anisotropy presented in Figs. 2-4, in Table 1 we present the average impedance mismatch scattering reflection coefficient for each of the three materials.  
As expected, diamond has by far the lowest impedance reflection coefficient, with an average R between one and two orders of magnitude less than PbTe and Bi$_2$Te$_3$.  The large values for PbTe suggest that, as with Bi$_2$Te$_3$, nanostructuring may yet be effective in reducing $\kappa_{lattice}$, if sufficiently small nanograins can be formed.  Conversely, the very small values of R for diamond provide a natural explanation for why polycrystalline diamond is such a good heat conductor.
\begin{table}[t!]
\caption{Average impedance mismatch reflection coefficients (multiplied by 100), and scattering potential coefficients, for three materials.  $L$ refers to the longitudinal mode and $T1$ and $T2$ to the transverse modes.}
\begin{center}
\begin{tabular}{|c|c|c|c|c|c|c|}
\hline
Compound  &  R$_{L}$  & R$_{T1}$ & R$_{T2}$ & R$_{total}$ & $\kappa_{lattice,bulk}$ (W/m-K)  & S$_{total}$ \\ \hline
Bi$_2$Te$_3$  &  0.153  & 0.429 & 0.437 & 0.340 & 1.7 & 1.16  \\ \hline
PbTe & 0.187 & 0.544 & 0.850 & 0.527 & 2.3 & 2.42 \\ \hline
diamond & 0.00627 & 0.0151 & 0.00537 & 0.0089 & 2200 &39.16\ \\ \hline
\end{tabular}
\end{center}
\end{table}

Here we have only considered sound velocity anisotropy, and in a real nanostructured sample numerous other factors will affect phononic transport, including (for example) the means of sample preparation, grain size and other microstructural properties.  In addition, other forms of scattering, such as by soft interstitial material between grains, could be a significant contributor to reducing thermal transport.  We do think, however, that the impedance and velocity mismatch associated with grain boundary scattering in nanostructured samples, as depicted here, can be a significant contributor to the reduction of thermal conductivity by nanostructuring, and that the quantitative parameters presented may give an indication of the likely effectiveness of nanostructuring in reducing the lattice thermal conductivity of a given material.

\section{Summary and Conclusions}

We present calculated phonon dispersions for \BiTe and discuss ceramic grain boundary scattering in terms of acoustic impedance mismatch.  We find that as expected grain boundaries
necessarily lead to strong interface scattering in \BiTe nanostructured material.  Interestingly, this is also expected to be the case in materials such as PbTe, which although cubic 
does have substantial acoustic wave anisotropy.  This is in contrast to the optical case, where a cubic material would have no such scattering.  In any case, the implication
of the present results is that dense sintered ceramics of anisotropic material such as \BiTe or PbTe will have reduced thermal conductivity provided
that the appropriate grain size is used.  In the case of more isotropic materials other strategies for producing scattering at grain boundaries, such as the introduction
of second phases, may be needed.
\\
{\bf Acknowledgments} 

This research was supported by the U.S. Department of Energy, EERE, Vehicle Technologies, Propulsion Materials Program (DP) and the ÔSolid State Solar-Thermal Energy Conversion Center (S3 TEC)Õ, an Energy Frontier Research Center funded by the US Department of Energy, Office of Science, Office of Basic Energy Sciences under Award Number: DE-SC0001299/DE-FG02-09ER46577 (DJS).

\end{document}